\documentclass[seceq]{ptptex}

\usepackage{epsfig}

\notypesetlogo    
\title{Overcoming system-size limitations in spin glasses}

\author{
Helmut G.~\textsc{Katzgraber}$^1$, 
M.~\textsc{K\"orner}$^1$, 
F.~\textsc{Liers}$^2$, and
A.~K.~\textsc{Hartmann}$^3$
}
\inst{
$^1$Theoretische Physik, ETH H\"onggerberg, CH-8093 Z\"urich, Switzerland \\
$^2$Institut f\"ur Informatik, Universit\"at zu K\"oln, Pohligstr. 1, D-50969 K\"oln, Germany \\
$^3$Institut f\"ur Theoretische Physik, Universit\"at G\"ottingen, Tammannstrasse 1, 37077 G\"{o}ttingen, Germany\\
}

\abst{
In order to overcome the limitations of small system sizes in spin-glass
simulations, we investigate the one-dimensional Ising spin chain with
power-law interactions. The model has the advantage over traditional
higher-dimensional Hamiltonians in that a large range of system sizes can
be studied.  In addition, the universality class of the model can be changed 
by tuning the power law exponent, thus allowing us to scan 
from the mean-field to long-range and short-range universality classes. We
illustrate the advantages of this model by studying the nature of the spin
glass state where our results hint towards a replica symmetry breaking
scenario. We also compute ground-state energy distributions and show that
mean-field and non-mean-field models are intrinsically different.
}

\begin{document}

\maketitle

\section{Introduction}
\label{sec:introdution}

Because spin glasses generally equilibrate very slowly due to their glassy 
behavior, simulations can usually only be performed 
on small system sizes. The introduction of
novel algorithms, such as exchange Monte Carlo (parallel
tempering),\cite{hukushima:96,marinari:98b} have helped considerably to
overcome these system-size limitations. Nevertheless, in general, only
modest system sizes can be simulated at temperatures well below the
critical temperature. In what follows we introduce the one-dimensional
Ising chain with random power-law interactions in order to simulate large
length scales. We illustrate the advantages of this model with two 
applications: the nature of the spin-glass state and ground-state energy 
distributions in spin glasses.

\section{Model \& Numerical Method}
\label{sec:model}

The Hamiltonian for the one-dimensional (1D) long-range Ising spin glass
with power-law interactions is given by\cite{bray:86b,fisher:88,katzgraber:03}
\begin{equation}
{\cal H} = -\sum_{i,j} J_{ij} S_i S_j\; , \;\;\;\;\;\;\;\;
J_{ij} \sim  \frac{\epsilon_{ij}}{r_{ij}^\sigma}\; , \;\;\;\;\;\;\;\;
{\mathcal P}(\epsilon_{ij}) = \frac{1}{\sqrt{2\pi}}\exp(-\epsilon_{ij}^2/2) 
\; ,
\label{eq:hamiltonian}
\end{equation}
where the sites $i$ lie on a ring of length $L$ to ensure periodic
boundary conditions, and $S_i = \pm 1$ represent Ising spins. The sum is
over all spins on the chain and the couplings $J_{ij}$ are Gaussian
distributed, and divided by the geometric distance between the spins to 
a power $\sigma$. The model has a rich phase diagram in the $d$--$\sigma$
plane:\cite{katzgraber:03} For $\sigma < 0.5$ the model is in the
mean-field phase with an SK universality class, for $0.5 \le \sigma \le
1.0$ the model has long-range critical exponents with a finite transition
temperature $T_{\rm c}$, whereas $T_{\rm c} = 0$ for $1.0 < \sigma
\lesssim 1.75$. For $\sigma \gtrsim 1.75$, again $T_{\rm c} = 0$ but in a
short-range universality class. In addition, there is a prediction from
droplet arguments\cite{bray:86b,fisher:88} that $\theta = d - \sigma$,
where $\theta$ is the stiffness exponent for domain-wall excitations.

For all of our simulations, we use the parallel tempering Monte Carlo
method\cite{hukushima:96,marinari:98b} as it allows us to study larger
systems (up to $L = 512$) at very low temperatures.  Details about the
simulations as well as equilibration tests can be found in
Ref.~\citen{katzgraber:03,katzgraber:03f}.

\section{Applications}
\label{sec:applications}

\subsection{Nature of the spin-glass state}
\label{subsec:nature}

Two main theories attempt to describe the nature of the spin-glass state:
The replica symmetry-breaking (RSB) picture and the ``droplet picture''
(DP). RSB predicts that droplet excitations involving a finite fraction of the
spins cost only a finite energy in the thermodynamic limit, and that the
surface of these excitations has a fractal dimension $d_s$ equal to the
space dimension $d$. In the DP, excitations have an energy proportional to
$\ell^{\theta}$, where $\ell$ is the characteristic length scale of the
droplet and $\theta$ is a positive stiffness exponent. In addition,
the excitation surfaces are fractal with $d_s < d$.\footnote{Recently Krzakala
and Martin, as well as Palassini and Young, suggest an intermediate
picture in which droplets have a fractal surface, and their energy is
finite in the thermodynamic limit. See Ref.~\citen{katzgraber:03} and
references therein for details.}. Differences between the various pictures
can be quantified by studying\cite{marinari:00,katzgraber:01} $P(q)$, the
distribution of the spin-glass overlap $q = L^{-1}\sum_{i = 1}^L
S_i^{\alpha} S_i^{\beta}$, where ``$\alpha$'' and ``$\beta$'' refer to two
replicas of the system with the same disorder. The RSB picture predicts a
nontrivial distribution with a finite weight in the tail around $q = 0$,
independent of system size. In contrast, the droplet picture predicts
that $P(q)$ is trivial in the thermodynamic limit with only two peaks at
$\pm q_{\rm EA}$, where $q_{\rm EA}$ is the Edwards-Anderson order
parameter. For finite systems there is also a tail down to $q = 0$, 
which vanishes in the thermodynamic limit like\cite{fisher:86,moore:98}
$P(0) \sim L^{-\theta'}$ with $\theta' = \theta$\footnote{Note that we
explicitly distinguish between $\theta^\prime$, the stiffness exponent for
droplet excitations, and $\theta$, the exponent for domain walls, because
DP predicts $\theta = \theta^\prime$, which we want to test here.}.

\begin{figure}[htbp]
\centerline{
    \includegraphics[width=6.5cm]{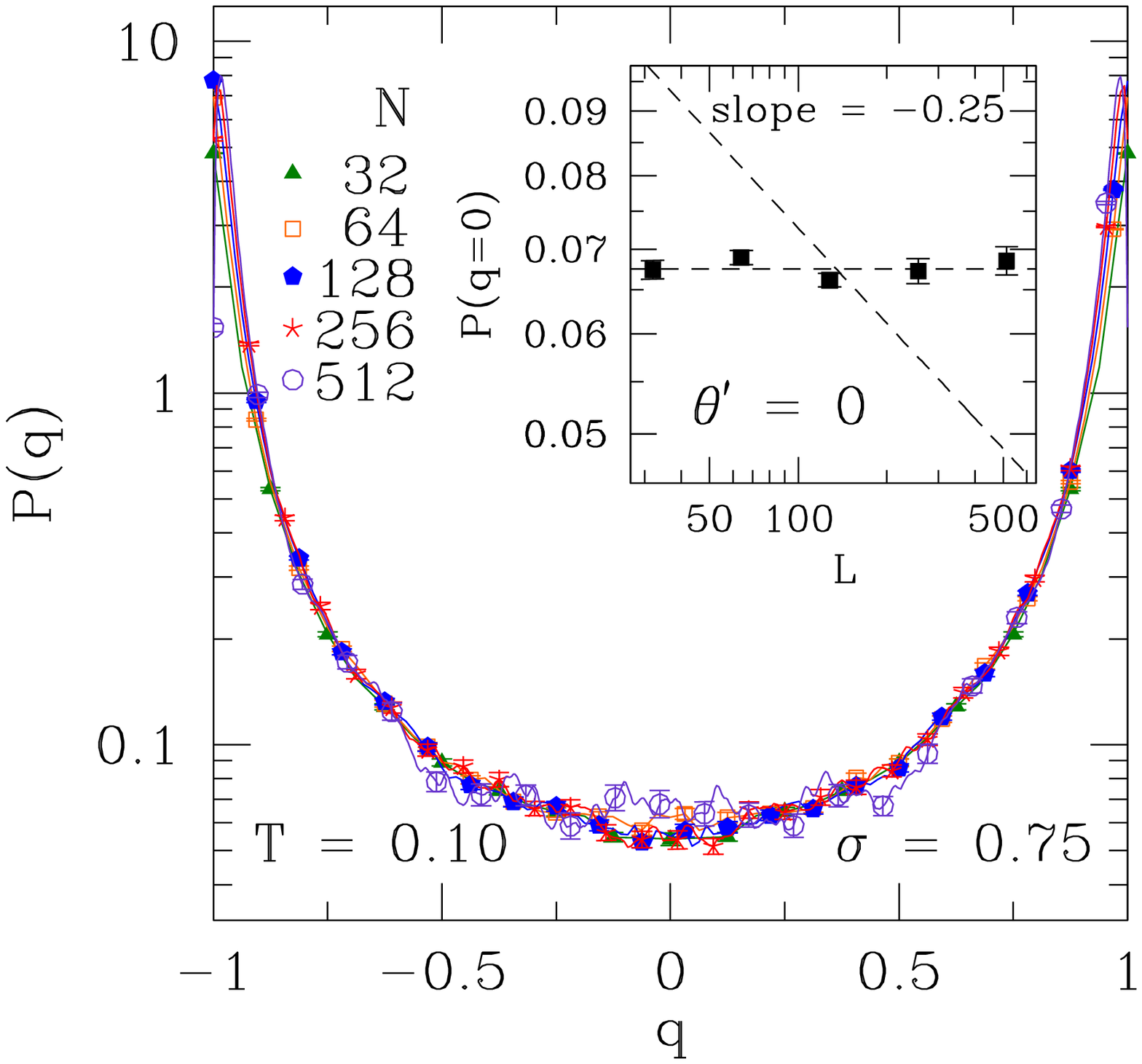}
    \includegraphics[width=6.5cm]{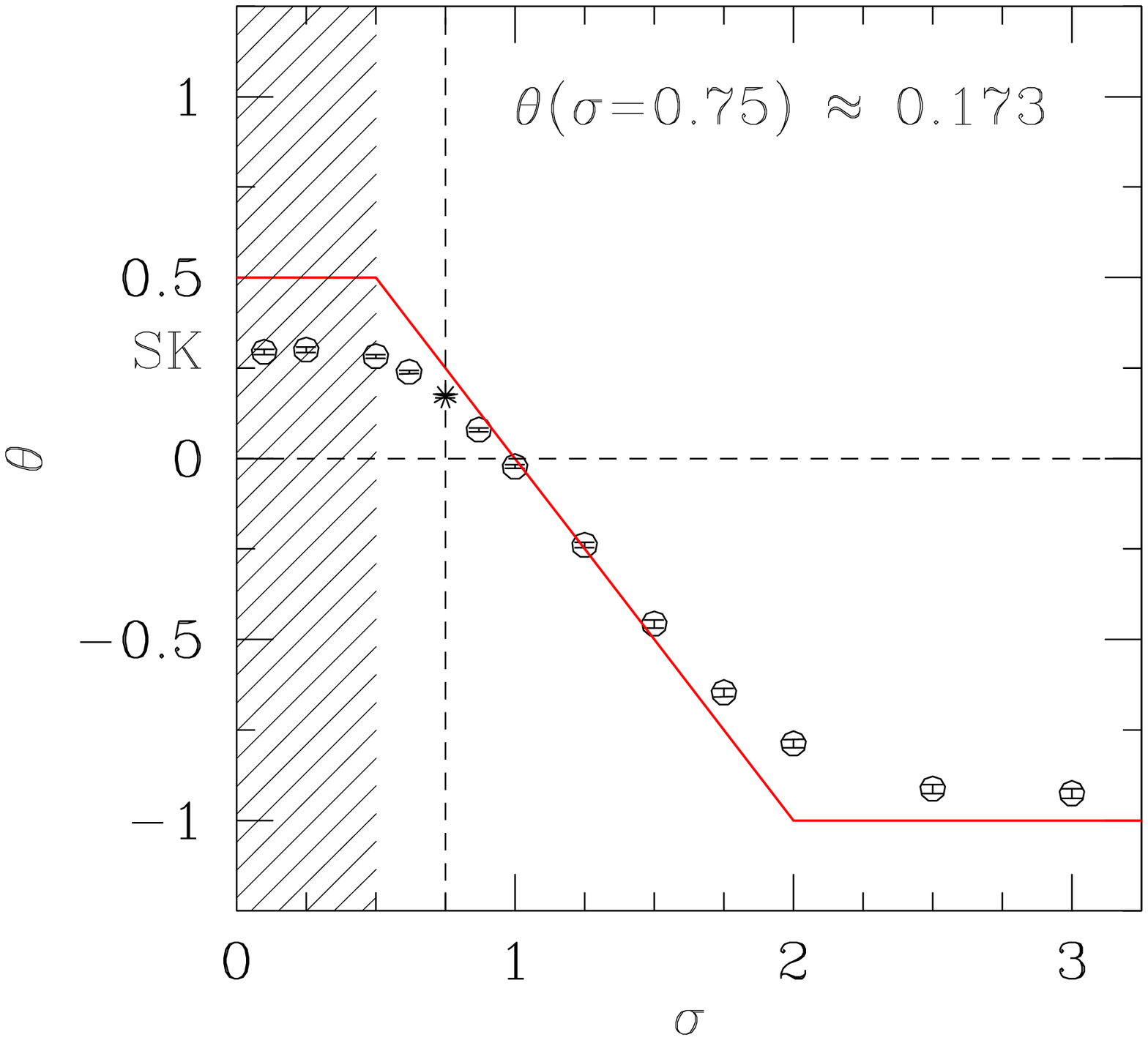}
    \vspace*{-0.5cm}
}
\caption{
Left panel: Distribution of the spin overlap $P(q)$ at $T = 0.10$ and
$\sigma = 0.75$ ($T_{\rm c}$ finite, long-range universality class) for
several system sizes $L$. The data are independent of $L$ at $q \sim 0$,
indicating that $\theta^\prime = 0$.  This is shown in detail in the inset
where $P(0) \sim L^{-\theta^\prime}$ is plotted as a function of $L$
together with the prediction from the droplet model (dashed line with
slope $-\theta = -(1 - \sigma) = -0.25$).  Right panel: stiffness exponent
$\theta$ as a function of $\sigma$ at $T = 0$ from domain-wall
calculations. The data follow well the prediction of the droplet picture
$\theta = d - \sigma$ (solid line). The star (``$\ast$'') marks $\sigma =
0.75$, where $\theta = 0.173 \pm 0.005$, which disagrees with
$\theta^\prime = 0$ from finite-temperature calculations (left panel).  
The dashed lines are guides to the eye. Note also that $\theta
\rightarrow 0.3$ for $\sigma \rightarrow 0$, the Sherrington-Kirkpatrick
(SK) model limit.
}
\label{fig:nature}
\end{figure}

Figure \ref{fig:nature}, left panel, shows data for $P(q)$ at $\sigma =
0.75$ and $T = 0.10$, well below $T_{\rm c} \approx
0.62$.\cite{katzgraber:03,leuzzi:99} There is a peak for large $q$ and a
tail down to $q=0$ that is independent of system size. A more precise
determination of the size dependence of $P(0)$ is shown in the inset of
Fig.~\ref{fig:nature}, left panel, where, to improve statistics, we
average over $q$ values with $|q| < 0.50$. The expected behavior in the
droplet model is $P(0) \sim L^{-\theta'}$, with $\theta' = \theta$
where\cite{bray:86b,fisher:88} $\theta = d - \sigma$. The dashed line has
slope $-0.25$, the expected value for $\sigma = 0.75$ according to the
droplet model.  The size dependence is consistent with a constant $P(0)$,
which implies that the energy to create a large excitation does not
increase with size, at least for the range of sizes studied here. In
Ref.~\citen{katzgraber:03f} we estimate the fractal dimension $d_s$ of the
system-size excitation surfaces and we find that $d_s \ge 0.95 \pm 0.05$,
i.e., $d_s \approx d$, in agreement with RSB.

Since there are analytic predictions for the stiffness exponent of the 1D
chain ($\theta = d - \sigma$), and the droplet picture predicts $\theta =
\theta^\prime$, we compute $\theta$ directly from zero-temperature
domain-wall calculations for different values of $\sigma$ using parallel
tempering as an optimization
algorithm.\cite{katzgraber:03,katzgraber:03f,hartmann:04}
The change in energy $\Delta E$ induced by a change in boundary conditions
from periodic (P) to antiperiodic (AP) scales as $\Delta E = [\,|E_{\rm
AP} - E_{\rm P}|\,]_{\rm av} \sim L^\theta$, where $[\cdots]_{\rm av}$
represents a disorder average. Data for $\theta(\sigma)$ are shown in
Fig.~\ref{fig:nature}, right panel. The solid line represents the
prediction from the droplet model, well followed by the data. For $\sigma
= 0.75$ (``$\ast$'' in Fig.~\ref{fig:nature}) there is a clear positive
stiffness exponent: $\theta(\sigma = 0.75) = 0.173 \pm 0.005$. This is in
contrast to $\theta^\prime = 0$ from finite-temperature simulations
(Fig.~\ref{fig:nature}, left panel). Therefore we find a disagreement with
the droplet model in that $\theta' \neq \theta$ for {\em a large range of
sizes}.

\subsection{Ground-state energy distributions}
\label{subsec:gs}

There has been recent interest in how ground-state energy distributions
scale in spin glasses. Considerable work has been done for the
Sherrington-Kirkpatrick (SK) model \cite{palassini:03a} where evidence for
skewed energy distributions, well fitted by a modified Gumbel
distribution\cite{bramwell:02} with $m = 6$, is found. In order to test if
these results are intrinsic to the mean-field SK model we compute
ground-state energy distributions for the 1D chain for different system
sizes and different values of $\sigma$. The advantage of the 1D chain is
that the crossover from mean-field to short-range behavior can be probed
for {\em a large range of sizes}.

\begin{figure}[htbp]
\centerline{
    \includegraphics[width=6.5cm]{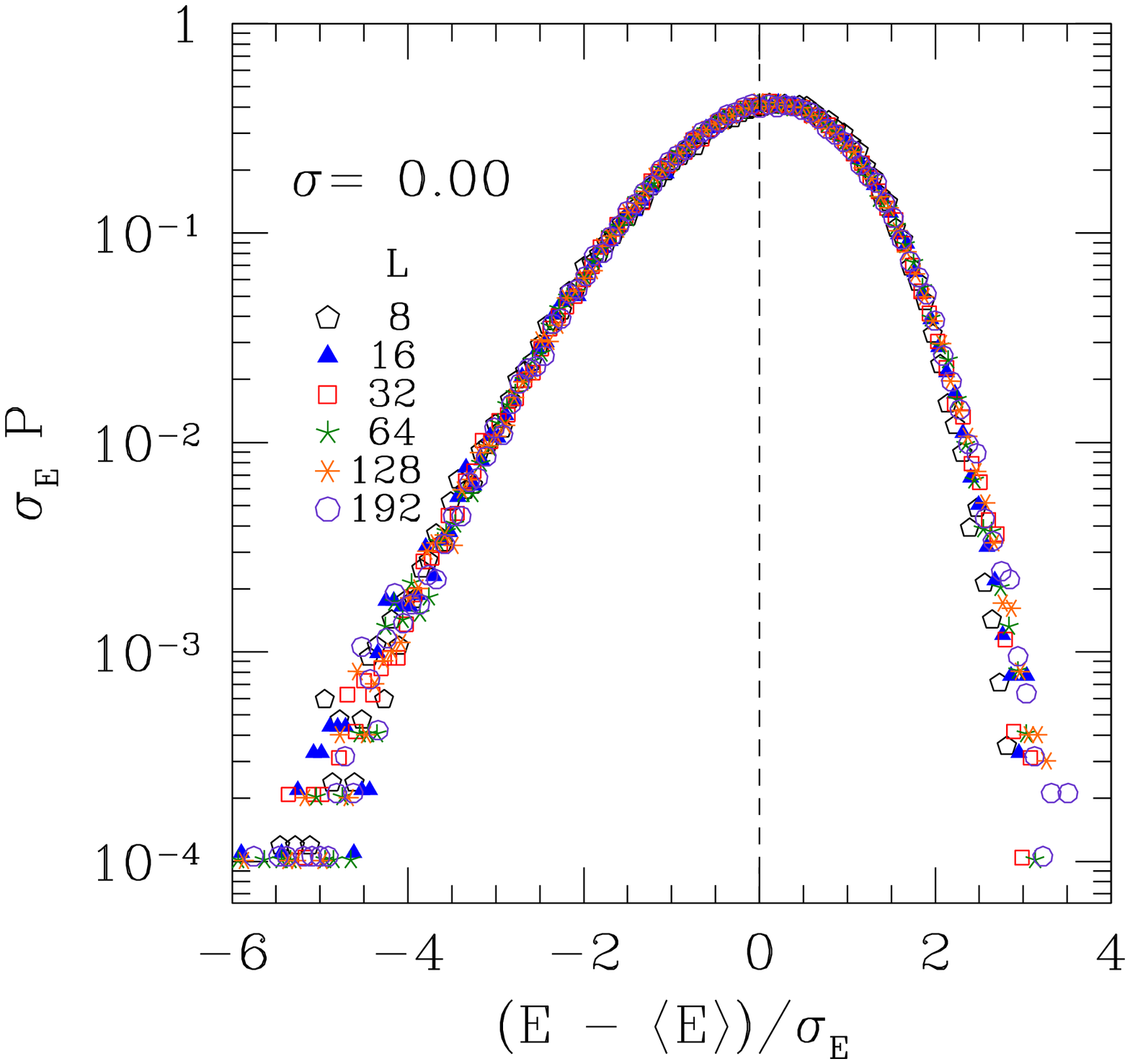}
    \includegraphics[width=6.5cm]{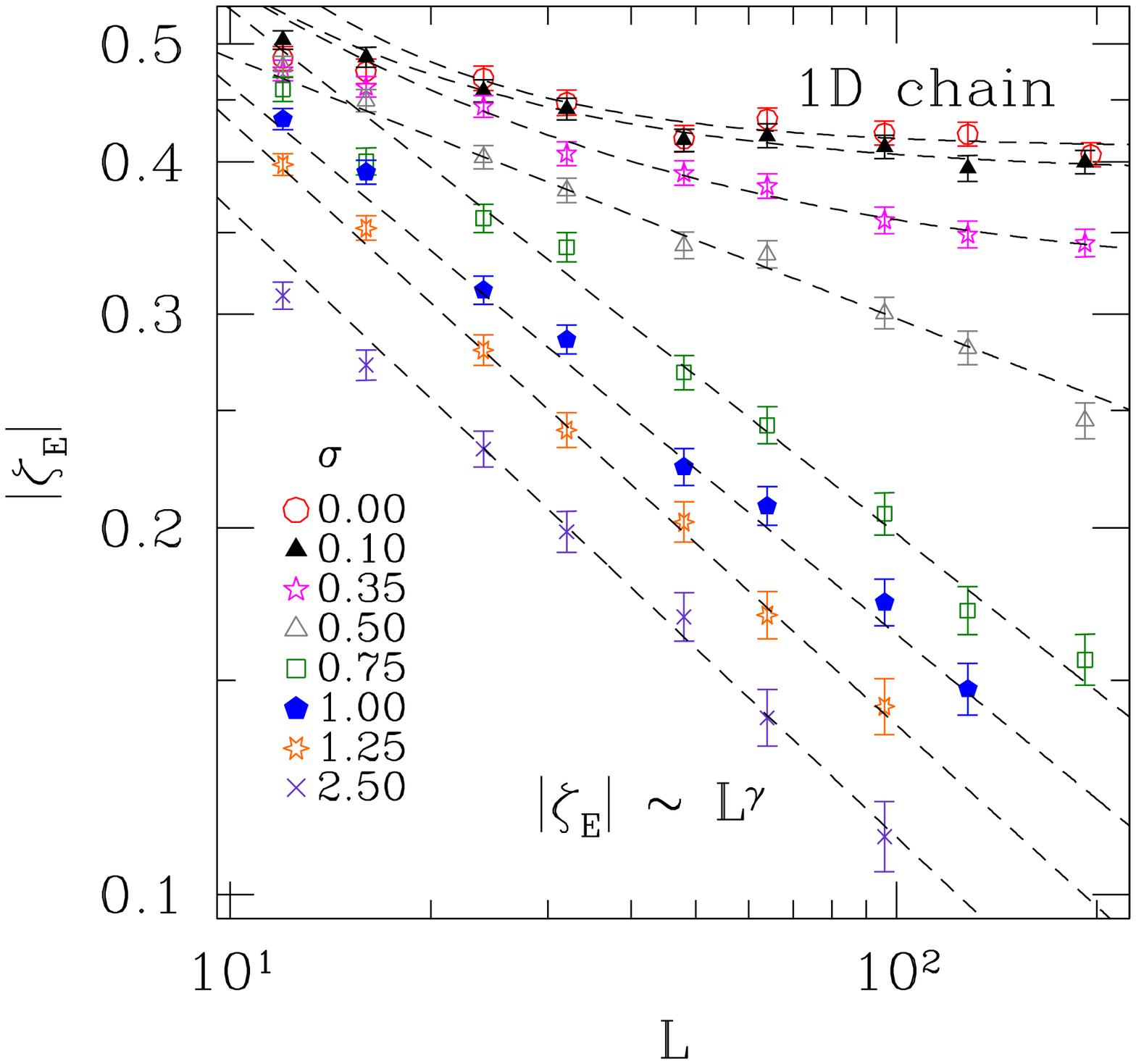}
    \vspace*{-0.5cm}
}
\caption{
Left panel: Rescaled ground-state energy distributions for $\sigma = 0.00$
(SK limit) as a function of system size $L$ ($10^5$ samples).  The data
are clearly asymmetric; the dashed line is a guide to the eye. Right panel:
Skewness $\zeta_{E}$ as a function of $L$ for different values of
$\sigma$. In the mean-field region ($\sigma < 0.5$) the $\zeta_{E}
\rightarrow {\rm const.}$ for $L \rightarrow \infty$. For $\sigma \ge 0.5$
the distributions become symmetric in the thermodynamic limit ($\zeta_{E}
\sim L^{\gamma}$, $\gamma < 0$).
}
\label{fig:gs}
\end{figure}

In Fig.~\ref{fig:gs}, left panel, we show rescaled ground-state energy
distributions $P(E)$ with mean $\langle E \rangle$ and standard deviation
$\sigma_{E}$. The data show a clear asymmetry. In order to quantify these
effects we show data for the skewness $\zeta_{E}$ of the distributions as 
a function of system size $L$ for different values of the power-law 
exponent $\sigma$ in Fig.~\ref{fig:gs}, right panel. One can see that for
$\sigma < 0.5$, where the model exhibits mean-field behavior, the skewness
tends to a constant in the thermodynamic limit, indicating that the
skewness of the distribution persists even for infinitely large system
sizes. For $\sigma \ge 0.5$, the skewness decays with a power law of the
system size, indicating that in the non-mean field region the ground-state
energy distributions become symmetric in the thermodynamic limit. This
shows that mean-field and non-mean field models are expected to behave
differently, and that, in particular, intrinsic length scales in the
mean-field model seem to scale with system size.

\section{Conclusions}
\label{sec:conclusions}

By using a one-dimensional Ising spin glass with power law interactions, we
have been able to overcome the usual limitation of small system sizes in
spin-glass simulations. We illustrate the advantages of the model on the
nature of the spin-glass state where we show that droplet and domain-wall
excitations scale differently for large system sizes, indicating that the
droplet model cannot be correct, at least for the 1D chain. In addition,
we compute the ground-state energy distributions for the one-dimensional
Ising chain and show that, in the mean-field regime, energy distributions
remain skewed in the thermodynamic limit indicating that mean-field and
non-mean-field models behave differently. Our results are in agreement with
previous results for the SK model ($\sigma = 0$). In the future, we intend
to study energy distributions at finite temperature, as well as the
effects of degenerate ground states.

\section*{Acknowledgments}
H.~G.~K.~would like to thank A.~P.~Young for a fruitful collaboration.
A.~K.~H.~was supported  by the {\em VolkswagenStiftung} (Germany) within the
program ``Nachwuchsgruppen an Universit\"aten''. Part of the simulations 
were performed on the Asgard cluster at ETH Z\"urich.

\bibliography{refs}

\end{document}